\documentclass[journal,12pt,onecolumn,draftclsnofoot,]{IEEEtran}
\usepackage{cite}
\usepackage{amsmath,amssymb,amsfonts}
\usepackage{algorithmic}
\usepackage{graphicx}
\usepackage{textcomp}
\usepackage{subcaption}
\usepackage{color,soul}
\usepackage{amsmath,amsfonts,amssymb,amsthm,epsfig,epstopdf,titling,url,array}
\usepackage{enumitem}

\theoremstyle{plain}
\newtheorem{thm}{Theorem}
\newtheorem{lem}{Lemma}

\theoremstyle{definition}
\newtheorem{defn}{Definition}

\theoremstyle{remark}
\newtheorem*{rem}{Remark}

\def\BibTeX{{\rm B\kern-.05em{\sc i\kern-.025em b}\kern-.08em
    T\kern-.1667em\lower.7ex\hbox{E}\kern-.125emX}}
\begin{document}
\title{Platooning in the Presence of a Speed Drop: A Generalized Control Model}

\author{Sina Arefizadeh\thanks{Sina Arefizadeh is a Ph.D Candidate in the Zachry Department of Civil Engineering at Texas A\&M University, College Station, TX 77843 USA (e-mail: sinaarefizadeh@tamu.edu). }~,
Alireza Talebpour\thanks{Alireza Talebpour, is an Assistant Professor in the Zachry Department of Civil Engineering at Texas A\&M University, College Station, TX 77843 USA. (e-mail: atalebpour@tamu.edu).}~, and
Igor Zelenko\thanks{Igor Zelenko is an Associate Professor in the Department of Mathematics at Texas A\&M University, College Station, TX
77843 USA, (e-mail: zelenko@math.tamu.edu).}}

\date{}
\maketitle

\begin{abstract}
Platooning is expected to enhance the efficiency of operating autonomous vehicles. The positive impacts of platooning on travel time reliability, congestion, emissions, and energy consumption has been shown for homogeneous roadway segments. However, unveiling the full potential of platooning requires stable platoons throughout the transportation system (end-to-end platooning). Speed limit changes frequently throughout the transportation network, due to either safety related considerations (e.g., change in roadway geometry and workzone operations) or congestion management schemes (e.g., speed harmonization systems). These abrupt changes in speed limit can result in shockwave formation and cause travel time unreliability. Therefore, designing a platooning strategy for tracking a reference velocity profile is critical to enabling end-to-end platooning. Accordingly, this study introduces a generalized control model to track a desired velocity profile, while ensuring safety in the platoon of autonomous vehicles. We  define appropriate natural error terms and the target curve in the state space of the control system, which is the set of points where all error terms vanish and corresponds to the case when all vehicles move with the desired velocities and in the miniml safe distance between them.  In this way we change the tracking velocity profile problem into a state-feedback stabilization problem with respect to the target curve. Under certain mild assumptions on the Lipschitz constant of the speed drop profile, we show that the stabilizing feedback can be obtained via introducing a natural dynamics for the maximum of the error terms for each vehicle. Moreover we show that with this stabilizing feedback collisions will not occur, if the initial state of the system of vehicles is sufficiently close  to the target curve. We also  show that the error terms remain bounded throughout the time and space.
 Two scenarios were simulated, with and without initial perturbations, and results confirmed the effectiveness of the proposed control model in tracking the speed drop, while ensuring safety and string stability.
\end{abstract}

\section{Introduction}
\label{sec:introduction}
\IEEEPARstart{C}{ombined} safety and efficiency of the transportation system has long been a topic of interest among traffic engineers. Human drivers tend to maximize their utility (e.g., minimize travel time and maximize safety). From the decision-making perspective, that means drivers tend to drive as close as possible to the free-flow speed, while keeping a safe distance from other drivers in the surrounding (particularly, the immediate leader and follower). Human factors, however, can cause uncertainties in this decision-making process and can potentially result in safety issues and congestion. Consequently, variability between expected travel time and the actual travel time is expected to increase, which results in reduction in travel time reliability. Note that any unreliability  and fluctuation in travel time can potentially result in user frustration, reduction in travel comfort, and further safety issues \cite{R1}.

Travel time of each vehicle is proportional to the inverse of its velocity. Therefore, tracking a certain velocity profile can guarantee a fixed amount of travel time and can maximize the reliability in transportation systems. In general, changes in speed limit are due to either safety considerations or changes in the roadway classification. For instance, a decrease in speed limit in workzone locations ensures drivers' responsiveness in dealing with unexpected situations. In a human dominated transportation system, tracking a sudden drop in the velocity profile (a change in speed limit) may produce backward waves in the traffic flow that can propagate upstream, change the traffic flow regime, and potentially result in flow breakdown \cite{helbing2009theoretical}.

Autonomous vehicles can potentially solve the above problem by removing the human element from driving related decisions \cite{talebpour2016influence}. Although the current state-of-the-practice in designing and testing autonomous vehicles is focused on isolated vehicles, it was shown that the effects of these vehicles on congestion and safety can improve by platooning. Implementing such a system requires additional infrastructure to provide Vehicle-to-Vehicle (V2V) and Vehicle-to-Infrastructure (V2I) communications \cite{R2}. Unfortunately, complete information about the system does not always exist due to signal interference and packet loss \cite{talebpour2016modeling}. One potential solution relies upon the combination of V2V/V2I communications and onboard sensros (e.g., radar and LiDAR sensors). Accordingly, several decentralized inter-connected platooning strategies have been proposed in the literature \cite{R8}. These studies consider two different platooning policies: constant spacing and constant time headway. The constant spacing policy is focused on keeping a constant space between the vehicles \cite{s1999cons}, while the constant time headway policy is focused on keeping a fixed time headway between consecutive vehicles in a platoon \cite{s2001rev, R9,R10}. Most of these studies focused on disturbances in the platoon and utilized the concept of "string stability" to ensure that any disturbance will decay as it propagates upstream in the platoon \cite{R14}. A formal definition of string stability was provided by Swaroop and Hedrick \cite{R15}. A generalization of the definition was also proposed by Pant et al. \cite{R16}, which is equivalent to stability in higher spatial dimensions. Moreover, alternative definitions of string stability for linear systems were also presented in other studies \cite{R17,R18}.

In addition to investigating string stability in the presence of disturbances, there exists few studies that focused on the problem of tracking a velocity profile \cite{R13, R21}. The approach in these studies is based on feedback linearization. Note that these studies rely on the existence of detailed information about the system. Such information should be transmitted to all vehicles, which requires reliable vehicular communications.

The problem of tracking a velocity profile, while ensuring safety requires controlling two error terms per vehicle (e.g., errors in tracking the velocity profile and keeping the safe distance) with one control input (e.g., vehicle acceleration). In general, introducing a controller that can ensure asymptotic stability for such a system is challenging. Note that in the linear control context, such problem can be addressed after stabilizing and by using servo machines for a class of reference velocity profiles (see Linear Control Theory, Lemma 13.6\cite{R19}) and in nonlinear control, sliding mode control can be adopted to resolve the tracking problem\cite{R20}. In the current research, we take an alternative approach and  introduce appropriate error terms and the target curve in the state space of the control system. The target curve is the set of points corresponding to the case where all error terms are vanished and all vehicles move at their desired speed, while keeping the minimum safe distance from the leader. Accordingly, the problem of tracking a velocity profile is changed into a state feedback stabilization problem with respect to the target curve. Such an approach can guarantee the minimum deviation from the velocity profile; thus, maximizes travel time reliability. Moreover, this study shows that such an approach results in string stability. 

Accordingly, the main contribution of this study is to introduce a novel methodology to deal with multiple outputs using one input, while ensuring asymptotic stability and a reasonable error bound in finite time (depending on the initial condition). To develop such a methodology, the notion of string stability for a target curve is defined. Utilizing an appropriate target curve, a time headway based state feedback is presented that ensures safety, while tracking the reference velocity profile for all vehicles in the platoon. Such a methodology can significantly reduce the level of difficulty of the platooning problem. In other words, using this method, the problem is simplified to the problem of finding a control law from a system of ordinary differential equations.

In addition to the above benefits, if vehicles are at their predefined location at the steady state condition (i.e., all vehicles are following the target curve), one can accurately predict the traffic flow regime. In other words, since all vehicles are at their reference points asymptotically, one can change the behavior of the system by changing the reference points. Therefore, in addition to the positive safety and congestion implications of the asymptotic behavior, designing/adjusting the asymptotic behavior of a string of vehicles provides the opportunity to further manage and regulate the traffic flow. As a result, specific densities and flows can be guaranteed at the steady state condition along with increasing the travel time reliability and ensuring safety.

The paper is organized as follows. First we discuss the main motivations toward introducing such an approach. In section \ref{sec:Problem Statement}, we introduce the target curve and stabilty with respect to it and  formulate the main result (Theorem \ref{theorem}) on state-feedback stabilization with respect to the target curve together with the treatment of noncollision and bounded control conditions. The proof of this theorem is given in Section \ref{sec:String Stability}. Section \ref{sec:Simulation Results} presents some simulation results and Section \ref{sec:Discussion} presents a discussion on the findings of this study. Finally, the paper is concluded with some remarks and future research needs in Section \ref{sec:Conclusion}.

\section{motivation}
\label{sec:motivation}
In the dynamic behavior of platoons, inter vehicular spacing plays a key role. Accordingly, two types of spacing policies are defined in the literature: constant spacing and constant time headway. The constant spacing maintains a fixed spacing between successive vehicles \cite{s1999cons}, while constant time headway operates with variable spacing defined based on a fix time headway \cite{s2001rev}. As discussed previously, some innovative time delayed strategies were also introduced in the literature \cite{R13, R21}.

The constant spacing policy intuitively implies that any changes in the velocity of a vehicle in the platoon should be transferred to the vehicles upstream (i.e., all the followers). Even though it can work for short platoons in the constant desired speed situations (e.g., Adaptive Cruise Control in a highway segment ), the definition of this spacing strategy makes it impossible to follow a speed profile defined in the spatial domain \cite{unpubmine}.  Accordingly, the constant time gap policy (also referred to as time delayed policy) was introduced \cite{R13}, \cite{R21}. Although it was shown that this policy can track a velocity profile \cite{R13, R21}, it brings its own challenges.

To clarify the difference between the constant time headway and the constant time gap policies, let us assume that the leader (vehicle $A$) passes a certain point, $M$, on the roadway segment at time $t_0$. The vehicle arrives at point $M'$ within time $G$ ($G$ stands for the time gap). Then the follower (vehicle $B$), which started from an arbitrary point at time $t_0$, reaches $M$ at $t_0+G$ with a given velocity. Figure \ref{fig0} illustrates the relative position of the vehicles with respect to the velocity profile at times $t_0$, $t_0+G$, and $t_0+NG$. Let $x _{i} (t), \dot{x} _{i} (t)$, and $v_d(x _{i} (t))$  denote the location, spot speed, and desired velocity of vehicle $i$, $i\in {\{A, B\}}$, at time $t$. Regarding the position of vehicles $A$ and $B$, we can write the following:
\begin{figure}
\includegraphics[width=9cm]{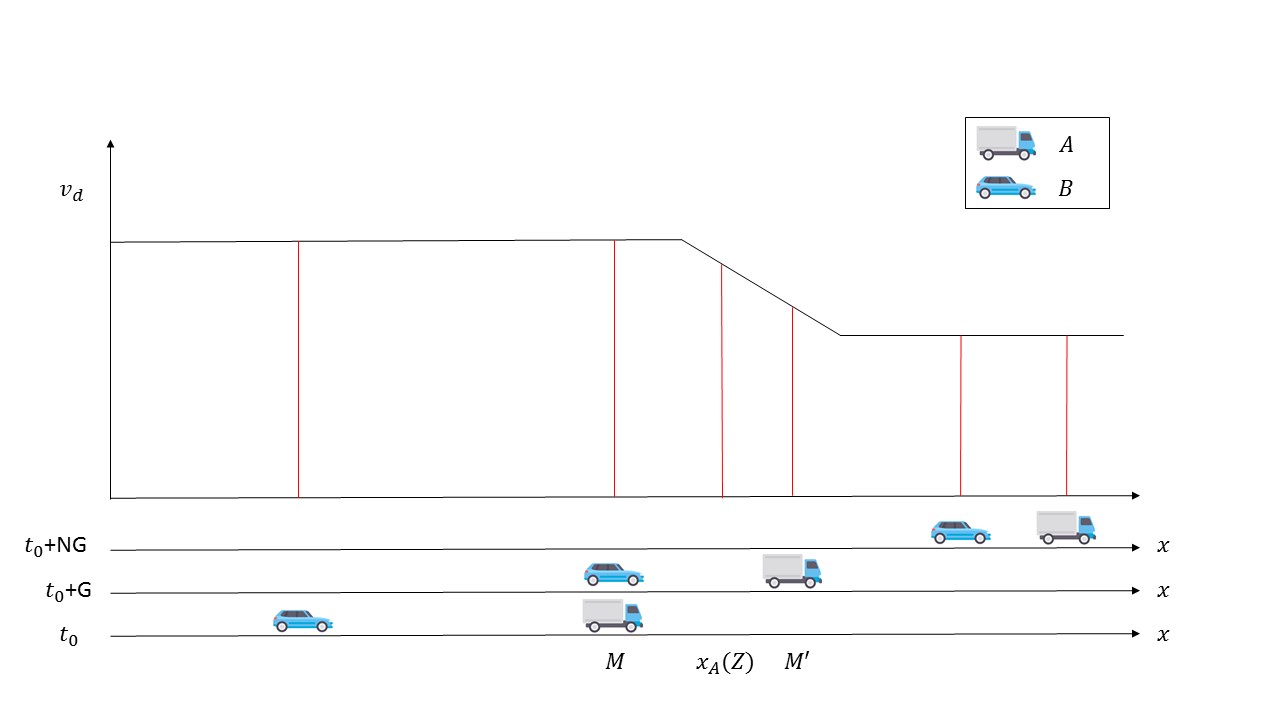}
\caption{Vehicle positioning with respect to velocity profile at $t_0$, $t_0+G$, and $t_0+NG$.}
\label{fig0}
\end{figure}

\begin{equation*}
\int^{t_0+G}_{t_0}\dot{x} _{A} (t)=x _{A} (t_0+G)-x _{B} (t_0+G).
\end{equation*}
In the ideal case where vehicle $A$ follows the velocity profile accurately:
\begin{equation*}
\dot{x} _{A} (t)=v_d(x _{A} (t)).
\end{equation*}
Therefore,
\begin{equation*}
\int^{t_0+G}_{t_0}v_d(x _{A} (t))=x _{A} (t_0+G)-x _{B} (t_0+G).
\end{equation*}
As a result, $\exists Z \in [t_0,t_0+G]$ such that:
\begin{equation*}
G.v_d(x _{A} (Z))=x _{A} (t_0+G)-x _{B} (t_0+G).
\end{equation*}
It should be noted that $x _{A} (Z) \in [x _{B} (t_0+G), x _{A} (t_0+G)]$. Clearly for the case that $v_d$ is not a constant velocity profile, $G$ does not yield $T$ considering the following:
\begin{equation*}
\begin{split}
G=\frac{x _{A} (t_0+G)-x _{B} (t_0+G)}{v_d(x _{A} (Z))}\ne\\
\frac{x _{A} (t_0+G)-x _{B} (t_0+G)}{\dot{x} _{B} (t_0+G)}.
\end{split}
\end{equation*}
The right handside of the above expression shows Time Headway of vehicle $B$ at the time $t_0+G$.
Therefore, in the fixed time gap policy, calculated time headways may not be constant. Thus, flow may vary throughout the platoon, which is not desirable and can result in forward or backward moving waves. Additionally, constant time gap cannot necessarily guarantee safety considerations \cite{unpubmine}. On the contrary, the definition of the constant time headway policy \cite{ioannou1993autonomous}, which is given below, shows that it ensures safety in the ideal case where there are no error terms in tracking the target location of vehicle $i$.

\begin{equation*}
x_{ref,i} (t) =x_{i-1} (t)-T\dot{x} _i (t),
\end{equation*}
where  $x_{ref,i} (t)$ is the ideal location of vehicle $i$, $x_{i-1} (t)$ denotes the actual location of vehicle $i-1$, and $\dot{x} _i (t)$ is the actual velocity of vehicle $i$ at time $t$. Based on this equation, if vehicle $i$ is at the ideal location, it takes $T$ seconds with its current speed to reach the position of vehicle $i-1$; therefore, it maintains a safe distance with the following vehicle. The above discussion motivates us to concentrate on the constant time headway policy and to introduce a control method to guarantee a fixed flow and safety based on this policy, even if vehicles are not initially at their predefined target locations (e.g., due to infrastructure heterogeneity).

\section{Problem Statement and Main Results}
\label{sec:Problem Statement}
\begin{figure}
\includegraphics[width=9cm]{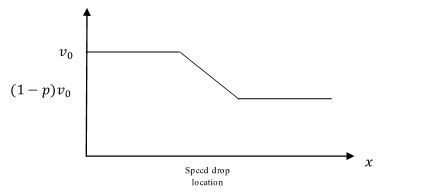}
\caption{Schematic Velocity Profile}
\label{fig1}
\end{figure}

Consider the desired velocity profile of Figure \ref{fig1}. This figure shows a gradual drop in the desired velocity. All vehicles are supposed to follow this velocity profile, where $v_{0}$ is the velocity of the string before the drop location, and $(1-\rho)v_{0}$  is the velocity of the string after the drop location. The quantity $\rho v_{0}$ represents the amount of velocity drop. Such velocity drops are common in transportation systems due to changes in speed limit. Our objective is to ensure that all vehicles in the platoon follow the profile presented in Figure \ref{fig1} and maintain a safe spacing from the immediate leader in the string over time and space, starting with a desirable initial condition (A discussion on the desirable initial condition is provided in the next section). Achieving this objective will maximize travel time reliability and ensure vehicle safety in the platoon.

Let us consider a cascaded system in which all plants has one available input and multiple error terms. For each plant, there is a vector of error terms of the same dimension for every plant. Let $\varepsilon_{ij} (t)$ denote the component $j ( j=1,..,m)$, of the error vector corresponding to plant $ i (i=1,\ldots,n)$ at time $t$ for all $m,n\in \mathbb{N}$. Two main assumptions are made in the current study:
\begin{itemize}
\item     The whole state of $(x_{i},\dot{x}_{i})$ can be measured directly so that the full state feedback is available without assigning and using an observer. A similar assumption was made in the study of longitudinal control of the lead car in a platoon by California PATH \cite{R22}.
\item  Following the study by Zheng et al. \cite{R23}, it is assumed that vehicle dynamics can be captured by a linearized model of vehicle. Note that a nonlinear model, considering drag, gravity, etc., is more realistic; however, it makes the problem more complex.
\end{itemize}

Ignoring the actuation lag, vehicle dynamics can be described by the platoon control system for all $n\in \mathbb{N}$ \cite{R24}

\begin{equation}
\label{second}
\ddot{x}_{i} (t)=u_{i}(t), \quad i=1,\ldots,n
\end{equation}
or, equivalently, by
\begin{equation}
\label{eq22}
\begin{cases}
            \dot{x} _i(t)=y_i(t)  \\
             \dot{y}_i(t)=u_i(t)
\end{cases}
\end{equation}


Define the following set on the phase space of $\mathcal X=\{(x_1,x_2,\ldots,x_n,y_1 ,y _2,\ldots,y_n): x_i, y_i\in \mathbb R\}$ of the system \eqref{eq22}:
\begin{equation*}
\begin{split}
\Gamma^{*}=\{x_1,x_2,\ldots,x_n,y _1,y_2,\ldots,y_n ):y _i=v_d (x_i),\\
            x_{j-1}=x_j+Ty_j,i=1,\ldots, n, j=2,\ldots, n\},
\end{split}
\end{equation*}
where $T$ denotes the time headway. Note that $\Gamma^*$ is a curve in  $\mathcal X$.
Indeed,
%
all components of the coordinate of a point  $(x_1,x_2,\ldots,x_n,y _1,y _2,\ldots,y _n)\in \Gamma^*$ are functions of $x_n$ only.

Since vehicles move with the desired velocity and the minimal safe distance on $\Gamma^{*}$, the following definition is natural.

\begin{defn}
We say that $\Gamma^{*}$ is \textit{the target curve}.
\label{def1}
\end{defn}

It will be an ideal situation, if  at every time moment the state of the system belongs to the target curve. The natural question is whether there exists a state feedback for the control system  \eqref{eq22} so that any initial configuration of vehicles will approach the target curve as $t\rightarrow+\infty$. To formalize what we mean by this, we give a series of definitions for stability of a curve, $\Gamma^*(t)$, with respect to a state-feedback. These definitions are  analogous to classical definitions of stability of critical points as well as the string stability concept defined in the transportation context \cite{R21}. In the remainder of this paper, the distance in the state space $\mathcal X$ is the Euclidean distance in  $\mathcal X$.

\begin{defn}
The target curve $\Gamma^*$ is \emph{stable with respect to the state feedback}
\begin{equation}
u_i=f\big(x_1(t),x_2(t),\ldots,x_n(t),\dot{x} _1(t),\dot{x} _2(t),\ldots,\dot{x} _n(t) \big),
\label{equation24}
\end{equation}
applied to system \eqref{eq22},
if for any $\epsilon>0$ there exists $\delta>0$
such that for all initial conditions
\begin{equation*}
x_i(0)=x_1^{0}, \dot{x}_i(0)=y_i^{0}, \quad i=1,\ldots, n,
\end{equation*}
with
\begin{equation}
 \label{IC}
\mathrm{dist}\Bigl(\bigl(x_1^0,x_2^0,\ldots,x_n^0,y_1^0,y_2^0,\ldots,y _n^0 \bigr),\Gamma^* \Bigl)<\delta
\end{equation}
the trajectory $\bigl(x_1(t), x_2(t), \ldots, x_n(t)\bigr)$
of the system
\begin{equation}
\label{dd}
\ddot{x} _i(t)= f\bigl(x_1(t),\ldots,x_n(t),\dot{x} _1(t),\ldots,\dot{x} _n(t)\bigr)
\end{equation}
satisfies
$$\mathrm{dist}\Bigl(\bigl(x_1(t),x_2(t),\ldots,x_n(t),\dot x_1(t),\dot x_2(t),\ldots,\dot x _n(t) \bigr),\Gamma^* \Bigl)<\varepsilon$$
for all $t>0$.
\label{defn20}
\end{defn}

\begin{defn}
The target curve $\Gamma^*$ is \emph{asymptotically stable with respect to the state feedback} \eqref{equation24}
applied to system \eqref{eq22} if it is stable in the sense of the previous definition and
there exists $\delta>0$ such that
for all initial conditions

\begin{equation*}
x_i(0)=x_1^{0}, \dot{x}_i(0)=y_i^{0}, \quad i=1,\ldots, n,
\end{equation*}
satisfying \eqref{IC}
the trajectory of the system \eqref{dd}
approaches $\Gamma^{*}$  as $t$ approaches $+\infty$, i.e.
\\$\mathrm{dist}\Bigl(\bigl(x_1(t),x_2(t),\ldots,x_n(t),\dot{x} _1(t),\dot{x} _2(t),\ldots,\dot{x} _n(t) \bigr),\Gamma^* \Bigl)$
tends to $0$ as $t\rightarrow+\infty$.
\label{defn2}
\end{defn}

\begin{defn}
 The curve $\Gamma^{*}$ is called \textit{globally asymptotically stable with respect to the state feedback \eqref{equation24}}
 if
any trajectory of the system \eqref{dd}
approaches $\Gamma^{*}$  as $t$ approaches $+\infty$.
\label{defn3}
\end{defn}

Note that in addition to providing asymptotic stability with respect to the target curve, the desired state feedback should prevent collisions and keep a reasonable bound for control inputs. Bounded control inputs are essential because engine power is limited and large values of control inputs are not practical.

The idea to construct the desired state feedback is based on  establishing the dynamics of the following error terms over time:
\begin{equation}
\varepsilon_{i1} (t)=\dot{x}_{i} (t)-v_{d} \big(x_{i} (t)\big)  \hspace{0.1cm}  for \hspace{0.1cm}   i=1,\ldots,n     ,
\label{eq3}
\end{equation}
\begin{equation}
\varepsilon_{i2} (t)=x_{i-1} (t)-x_{i} (t)-T\dot{x}_{i} (t) \hspace{0.1cm} for \hspace{0.1cm}  i=1.\ldots,n,
\label{eq4}
\end{equation}
By substituting the expression \eqref{eq3} and \eqref{eq4} for the error terms $\varepsilon_{i1}(t)$ and $\varepsilon_{i2}(t)$ into

\begin{equation}
\begin{cases}
             \dot{\varepsilon}_{i1} (t)=-\varepsilon_{i1} (t) & \text{if $ |\varepsilon_{i1} (t)|\geq|\varepsilon_{i2} (t)| $}\\
             \dot{\varepsilon}_{i2} (t)=-\varepsilon_{i2} (t) & \text{if $ |\varepsilon_{i1} (t)|<|\varepsilon_{i2} (t)|, $}
\end{cases}
\label{eq2}
\end{equation}
we  have the following system
\begin{equation}
  \begin{cases}
              \ddot{x}_{i} (t)=\dot{x}_{i} (t)v_{d}^{'} \big(x_{i} (t)\big)-\dot{x}_{i} (t)+v_{d} \big(x_{i} (t)\big)\\

\text{if $ |\dot{x}_{i} (t)-v_{d} \big(x_{i} (t)\big)|\geq|x_{i-1} (t)-x_{i} (t)-T\dot{x}_{i}(t)| $}\\
              \ddot{x}_{i} (t)=\frac{1}{T} \big(x_{i-1} (t)-x_{i} (t)-T\dot{x}_{i} (t)+\dot{x}_{i-1} (t)-\dot{x}_{i} (t)\big)\\

\text{if $ |\dot{x}_{i} (t)-v_{d} \big(x_{i} (t)\big)|<|x_{i-1} (t)-x_{i} (t)-T\dot{x}_{i} (t)| ),$}
\end{cases}
\label{eq5}
\end{equation}
for $i =1,\ldots,n$,  $n\in\mathbb{N}$.
This system is in fact from the control system \eqref{eq22} by applying the state feedback

\begin{equation}
    u_{i}=\begin{cases}
              \dot{x}_{i} (t)v_{d}^{'} \big(x_{i} (t)\big)-\dot{x}_{i} (t)+v_{d} \big(x_{i} (t)\big)\\

\text{if $ |\dot{x}_{i} (t)-v_{d} \big(x_{i} (t)\big)|\geq|x_{i-1} (t)-x_{i} (t)-T\dot{x}_{i}(t)| $}\\
              \frac{1}{T} \big(x_{i-1} (t)-x_{i} (t)-T\dot{x}_{i} (t)+\dot{x}_{i-1} (t)-\dot{x}_{i} (t)\big)\\

\text{if $ |\dot{x}_{i} (t)-v_{d} \big(x_{i} (t)\big)|<|x_{i-1} (t)-x_{i} (t)-T\dot{x}_{i} (t)| ).$}
\end{cases}
\label{eq1}
\end{equation}
The main results of the paper are gathered in the following theorem.



\begin{thm}
\label{theorem}
Assume that the speed drop profile $v_d$ is  Lipschitz with the Lipschitz constant $M<\frac{1}{T}$ , where $T$ stands for the constant time headway.
Then the following statements hold:

\begin{enumerate}
 \item
The target curve $\Gamma^*$ is globally asymptotically
stable with respect to the state feedback given by \eqref{eq1};

\item (non-collision conditions) If, in addition, the desired velocity profile $v_d$ satisfies $\inf\{v_d(x):x\in\mathbb R\}>0$ then for all initial conditions
\begin{equation*}
x_i(t_0)=x_i^{0}, \dot{x}_i(t_0)=y_i^{0}, \quad i=1,\ldots, n,
\end{equation*}
satisfying
\begin{equation}
\begin{split}
\label{maxerror}
\mathcal \mathrm{dist}(x_1^0,\ldots, x_n^0, y_1^0,\ldots, y_n^0,\Gamma^*)\\\leq\cfrac{T \inf\{v_d(x):x\in\mathbb R\}}{\max{\{2+T,1+M\}}(1+T)},
\end{split}
\end{equation}
the solution $\big(x_1(t),\ldots x_n(t)\big)$ of  the system \eqref{eq5}
satisfies $x_{i-1}(t)-x_i(t)>0$ for any $t\geq t_0$;

\item If, in addition to item 1), the speed profile $v_d$ is bounded and  piecewise differentiable  with bounded derivatives, then the control inputs in the  state feedback  \eqref{eq1} are bounded along each trajectory of \eqref{eq5} for $t\geq t_0$.
\end{enumerate}
\end{thm}

The proof of Theorem \ref{theorem} will be given in the next section. A controller of the form (\ref{eq1}) is a distributed controller that receives information about the velocity profile from infrastructure (i.e., V2I communications). The decentralized nature of this model eliminates the need for information from all the vehicles in the platoon. Such a requirement (receiving velocity and/or location information from every vehicle) can decrease reliability, safety, and efficiency of the entire system as signal interference can create information loss.

\begin{rem}
A similar result with almost the same proof is true  for more general state feedback that is obtained by using the following system of differential equations for error terms instead of \eqref{eq2}
\begin{equation*}
\begin{cases}
             \dot{\varepsilon}_{i1}  (t)=g_{i1}\big(\varepsilon_{1} (t)\big) & \text{if $ |\varepsilon_{1} (t)|\geq|\varepsilon_{2} (t)| $}\\
             \dot{\varepsilon}_{i2} (t)=g_{i2}\big(\varepsilon_{2} (t)\big) & \text{if $ |\varepsilon_{1} (t)|<|\varepsilon_{2} (t)|, $}
\end{cases}
\label{eq2*}
\end{equation*}
for $i=1,\ldots, n$,
where $g_{ij}(\cdot)$ are such that the equations $\dot{x}  (t)=g_{ij}\big(x (t)\big)$ are globally  asymptotically stable at the origin for $j=1,2$.
\end{rem}


\section{Proof of Theorem \ref{theorem}}
\label{sec:String Stability}
This section is focused on the proof of Theorem \ref{theorem}.
\subsection {Proof of item 1)}
This Proof is divided into two lemmas.
Lemma \ref{lem1} implies that under the state feedback \eqref{eq1}, the error terms \eqref{eq3} and \eqref{eq4} decay exponentially as $t\rightarrow +\infty$.
This result is independent of the assumption that $v_d$ is Lipschits with a specific upper bound for the Lipschitz constant.  Lemma 2 implies that under assumptions on the Lipschitz constant, as in Theorem \ref{theorem}, the decay of the error terms  \eqref{eq3} and \eqref{eq4} yields the decay of the distance between the trajectory of the system \eqref{eq5} and the target curve.


Since the following arguments are independent of $n$, for simplicity, we use $\varepsilon_{1}$ and $\varepsilon_{2}$ instead of $ \varepsilon_{i1}$ and $\varepsilon_{i2}$ in the system (\ref{eq2}).

\begin{lem}
\label{lem1}
For a system of differential equations of the form (\ref{eq2})
and $\delta(t)=max⁡\{|\varepsilon_{1} (t)|,|\varepsilon_{2} (t)|\}$, we have
 $\delta(t)=Ae^{-t}$, where $A=\delta(t_{0} ) e^{t_{0} }$.
\end{lem}

\textbf{Proof.}\hspace{0.1cm}  let $ S=\{t\in \Bbb R\colon|\varepsilon_{1} (t)|=|\varepsilon_{2} (t)|\},  S$ is a closed set.
\begin{enumerate}[label=(\roman*)]
\item On $\Bbb R\backslash S$ (the compliment of $S$ to $\Bbb R$), $\delta(t)=\pm\varepsilon_{i} (t)$ for some $ i\in\{1,2\}$. Therefore, $\forall t\in \Bbb R\backslash S, \dot{\delta} (t)=-\delta(t)$. As a result, $\delta(t)=Ae^{-t}$ on each connected component of $\Bbb R\backslash S$ (The maximal connected subsets of a nonempty topological space are called the connected components of the space).
\item On $Int(S)$ (interior of $ S$), $\delta(t)=\varepsilon_{1} (t)$. Thus, $\dot{\delta}(t)=-\delta(t)$. As a result, $\delta(t)=Be^{-t}$ on each connected component of $Int(S)$.
\item For $\forall t\in \partial S \hspace{0.1cm} (\partial S=S\backslash Int(S))$, we can use continuity of $ \delta(t)$ to conclude that $\delta(t)=Ae^{-t}$. From (i) and (ii) and continuity, one can conclude that constants near exponential terms are the same. This completes the proof.\hspace{0.1cm}$\blacksquare$
\end{enumerate}

Now define the following function $\mathcal E$ on the state space $\mathcal X$ of \eqref{eq22}
\begin{equation}
\begin{split}
\mathcal E(x_1,x_2,\ldots,x_n,y _1,y _2,\ldots,y_n)=\hspace{0.75cm}\\\max_{i,j\in{1,2,\ldots,n}}⁡\{|y_i-v_d (x_i)|,|x_{j-1}-x_j-Ty_j| \}.
\end{split}
\label{eq21}
\end{equation}
Note that the target curve $\Gamma^*$ is nothing but the zero level set of the function $\mathcal E$.

Take a point
$$Q=\bigl(x_1,\ldots, x_n, y_1\ldots, y_n\bigr)$$
in the state space $\mathcal X$. Then directly from the definition of the target curve  $\Gamma^*$, there exists the unique point
$$ Q^*=\bigl(x_1^*,\ldots, x_n^*, y_1^*,\ldots, y_n^*\bigr)$$
on $\Gamma^*$
such that $x_1^*=x_1$.


\begin{lem}
\label{lem4}
If the speed drop profile $v_d$ is  Lipschitz with the Lipschitz constant $M<\frac{1}{T}$,
then
\begin{equation}
|x_i-x_i^*|\leq C_i (1+T) \mathcal E(Q)\quad i>1,
\label{eq24}
\end{equation}
\begin{equation}
|y _i-y _i^*|\leq (C_i M(1+T)+1 )\mathcal E(Q)\quad  i\geq1
\label{eq25}
\end{equation}
where $$C_i=\sum_{k=2}^{i}\frac {1}{(1-TM)^{i-k+1} }.$$
Consequently, there exists a constant $C>0$ such that for any point $Q$ in the state space $\mathcal X$,
\begin{equation}
\label{distest}
\mathrm{dist}(Q, \Gamma^*)\leq C \, \mathcal E(Q).
\end{equation}
\end{lem}

\textbf{Proof.} Set
\begin{equation*}
\mu_i=x_i-x_i^*, \quad \eta_i=\dot{x} _i-\dot{x} _i^*.
\end{equation*}
Also, by analogy with \eqref{eq3} and \eqref{eq4}, let
\begin{eqnarray}
&~&\label{error1} \varepsilon_{i1}=y_i-v_d (x_i),\\
&~&\label{error2} \varepsilon_{i2}=x_{i-1}-x_i-Ty_i.
\end{eqnarray}
Let $\delta_i=\max\{\varepsilon_{i1},\varepsilon_{i2}\}$.
Then the following recursive inequalities hold
\begin{eqnarray}
&~&|\eta_i|\leq M|\mu_i|+|\varepsilon_{i1} |,\,\,\,\, i\leq  n, \label{eq39}\\
&~&|\mu_{i+1}|\leq\frac{1}{1-TM} |\mu_i|+\frac{1+T}{1-TM} \delta_{i+1},\,i< n.\label{eq38}
\end{eqnarray}
Based on \eqref{error1} and \eqref{error2}

\begin{equation*}
\begin{cases}
             y_{i+1}=v_d (x_{i+1})+\varepsilon_{i+1,1}\\
             x_{i+1}=x_i-T y_{i+1}-\varepsilon_{i+1,2}
\end{cases},
\end{equation*}
\begin{equation*}
\begin{cases}
             y _{i+1}^*=v_d (x_{i+1}^*)\\
             x_{i+1}^*=x_i^*-Tv_d (x_{i+1}^*)
\end{cases}.
\end{equation*}\\
Then first
\begin{equation}
\eta_{i+1}=v_d (x_{i+1})-v_d (x_{i+1}^*)+\varepsilon_{i+1,1}.
\label{eq35}
\end{equation}
Since $v_d$ is Lipschitzian with Lipschitz constant $M$ we get \eqref{eq39}.

Furthermore,
\begin{equation*}
\begin{split}
\mu_{i+1}=x_{i+1}-x_{i+1}^*\hspace{5cm}\\
=\mu_i-T\big(v_d (x_{i+1})-v_d (x_{i+1}^*) \big)-T\varepsilon_{i+1,1}-\varepsilon_{i+1,2},
\end{split}
\label{eq33}
\end{equation*}
Similar to the above case, since $v_d$ is Lipschitzian with Lipschitz constant $M$, we get
\begin{equation}
|\mu_{i+1}|\leq \mu_i+M T|\mu_{i+1}|+(1+T)\delta_{i+1},
\label{eq34}
\end{equation}
which together with the condition $MT<1$ imply the inequality \eqref{eq38}.

Applying the recursive formula as many times as necessary and using $\mu_1$=0, we obtain
\begin{equation}
|\mu_i|\leq(1+T) \sum_{k=2}^{i}\frac {\delta_k}{(1-TM)^{i-k+1}}
\label{eq40}
\end{equation}
which implies inequality \eqref{eq24}

This equation together with (\ref{eq39}) implies \eqref{eq25}. Finally, the inequality \eqref{distest} follows immediately from inequalities \eqref{eq39} and \eqref{eq38} and definition of the function $\mathcal E$.
$\blacksquare$\

Now, if $\Gamma(t)$ is a trajectory of system \eqref{eq5}, then from Lemma 1 it can be concluded that $\mathcal E\big(\Gamma(t)\big)\leq \mathcal E\big(\Gamma(t_0)\big)e^{t_0-t}$, which together with inequality \eqref{distest} implies that $\mathrm{dist} (\Gamma(t),\Gamma^*)\leq C \mathcal E\big(\Gamma(t_0)\big) e^{t_0-t}$. This completes the proof of item 1) of Theorem \ref{theorem}.

\subsection{Proof of item 2)}
%
%
Assume that
$$\Gamma(t)=\big(x_1(t),\ldots x_n(t), \dot x_1(t), \ldots, \dot x_n(t)\big).$$
Rewriting (\ref{eq4}) based on (\ref{eq3}) and using Lemma \ref{lem1} we get
\begin{equation}
\begin{split}
x_{i-1} (t)-x_i (t)= T v_d \big(x_i (t)\big)+T \varepsilon_{i1} (t)+\varepsilon_{i2} (t)\geq \\
T  \inf \{⁡v_d (x):x\in \mathbb R\}-(1+T)\mathcal E(Q^0).
\end{split}
\end{equation}
for $t\geq t_0$, where $Q^0$ denotes the initial condition. Hence, $x_{i-1} (t)-x_i (t)>0$ for all $t>0$, if
\begin{equation}
\label{prelimineq}
\mathcal E(Q^0)< \cfrac{T  \inf \{⁡v_d (x):x\in \mathbb R\}}{1+T}. \end{equation}

\begin{lem}
\label{lemma30}
For all $Q\in\mathcal X$  we have
\begin{equation}
\mathcal E(Q)\leq \max{\{2+T,1+M\}} \mathrm{dist}(Q,\Gamma^*).
\label{eqlemma30}
\end{equation}
\end{lem}

\textbf{Proof.} First, given a point $Q\in\mathcal X$, the closest point of $\Gamma^*$ to $Q$ exists. Indeed, $\Gamma^*$ is parameterized by $x_1$ and for a sufficiently large number $N>0$, the distance from any  points on $\Gamma^*$  with $|x_1|>N$  to the point $Q$ is at least twice of $\mathrm{dist}(Q,\Gamma^*)$. So the distance form $Q$ to $\Gamma^*$ is the same as the distance from $Q$ to a part of $\Gamma^*$ corresponding to $|x_1|\leq N$. The latter part of $\Gamma^*$ is compact so that the infimum of the distance from $Q$ to $\Gamma^*$ is indeed achieved at some point $Q^*$ of $\Gamma^*$.

Assume that $Q^*=(x_1^*,\ldots, x_n^*, y_1^*,\ldots, y_n^*)$. Set $\Delta:=\mathrm{dist}(Q,\Gamma^*)=\mathrm{dist}(Q,Q^*)$. To prove inequality \eqref{eqlemma30} we need first to estimate $|x_{i-1}-x_i-Ty_i|$. By triangle inequality it is less than or equal to
\begin{equation*}
|x_{i-1}-x_i-Ty_i^*|+T|y_i^*-y_i|,
\end{equation*}
for $i=1,\ldots,n$ for all $n\in \mathbb N$.

By assumptions,  
\begin{equation*}
\begin{cases}
            |x _i-x_i^*|\leq\Delta  \\
            |y _i-y_i^*|\leq\Delta
\end{cases}
for \hspace{0.1cm} i=1,\ldots,n.
\label{eq101}
\end{equation*}
Therefore,
\begin{equation}
\label{eq102}
\begin{split}
|x_{i-1}-x_i-Ty_i^*|=|x_{i-1}-x_i- (x_{i-1}^*-x_i^*)|\\\leq |-x_i+x_i^*|+|x_{i-1}-x_{i-1}^*|\leq 2\Delta.
\end{split}
\end{equation}
Considering (\ref{eq102}) we have
\begin{equation*}
\begin{split}
|x_{i-1}-x_i-Ty_i^*|+T|y_i^*-y_i|\\\leq (2+T)\Delta,
\end{split}
\end{equation*}
thus
\begin{equation}
|x_{i-1}-x_i-Ty_i|\leq(2+T)\Delta,
\label{eq103}
\end{equation}
for $i=1,\ldots,n$.

Further, estimate $|y_i-v_d(x_i)|$:
\begin{equation*}
|y_i-v_d(x_i)|\leq|y_i-y_i^*|+|y_i^*-v_d(x_i)|.
\end{equation*}
Since $v_d(.)$ is Lipschitz and $y_i^*=v_d(x_i^*)$ we have
\begin{equation*}
|y_i^*-v_d(x_i)|\leq M|x_i-x_i^*|,
\end{equation*}
then
\begin{equation}
\label{eq104}
|y_i-v_d(x_i)|\leq (1+M)\Delta
\end{equation}
for $i=1,\ldots,n$.
Then (\ref{eq103}) and (\ref{eq104}) result in the inequality (\ref{eqlemma30}).
$\blacksquare$

Combining inequality \eqref{prelimineq}  and Lemma \ref{lemma30}, we get that inequality \eqref{maxerror} implies that $x_{i-1} (t)-x_i (t)>0$, which completes the proof
of item 2) of Theorem \ref{theorem}.


\subsection{Proof of item 3)}

Let $\big(x_1(t),\ldots, x_n(t)\big)$ be a solution of \eqref{eq5}, $\varepsilon_{i1}(t)$ and $\varepsilon_{i2}(t)$ are as in \eqref{eq3} and \eqref{eq4}, and $\delta_i(t)=\max\{|\varepsilon_{i1}(t)|, |\varepsilon_{i2}(t)|\}$.
Rewriting control input in  (\ref{eq1}) for the case of $|\dot{x} _i (t)-v_d \big(x_i (t)\big)|\geq|x_{i-1} (t)-x_i (t)-T\dot{x} _i (t)|$, we have

\begin{equation*}
\centering
u_i=\Big(v_d \big(x_i (t)\big)+\varepsilon_{i1} (t)\Big)v_d^{'} \big(x_i (t)\big)-\varepsilon_{i1} (t).
\label{eq17}
\end{equation*}
Consequently, using Lemma \ref{lem1} we get that for all $t\geq t_0$
\begin{equation}
\centering
|u_i |\leq\big(|v_d \big(x_i (t)\big)|+\delta_i (t_0 )\big)|v_d^{'} \big(x_i (t)\big)|+\delta_i (t_0 ).
\label{eq18}
\end{equation}
For the case of $|\dot{x}_i (t)-v_d \big(x_i (t)\big)|<|x_{i-1} (t)-x_i (t)-T\dot{x}_i (t)|$, we have
\begin{equation*}
\begin{split}
u_i=\frac{1}{T} \bigg(\varepsilon_{i2} (t)+\Big(v_d \big(x_{i-1} (t)\big)+\varepsilon_{i-1,1} (t)\Big)-\\
\Big(v_d \big(x_i (t)\big)+\varepsilon_{i1} (t)\Big)\bigg).
\end{split}
\label{eq19}
\end{equation*}
Therefore, using Lemma \ref{lem1}, we have
\begin{equation}
\begin{split}
|u_i |\leq \frac{1}{T} \Big(|v_d \big(x_{i-1} (t)\big)|+|v_d \big(x_i (t)\big)|+\\ 2 \delta_i (t_0 )+\delta_{i-1} (t_0 )\Big)
\end{split}
\label{eq20}
\end{equation}
Combining estimates (\ref{eq18}) and (\ref{eq20}) with the assumption that $v_d$ is bounded and have piecewise continuous bounded derivative, we can conclude that the control input $u_i(t)$ is bounded for $t\geq t_0$. $\blacksquare$


\begin{rem}
The differential equation of the form (\ref{eq2}) switches between two linear systems: $\dot{\varepsilon}_{2}  (t)=-\varepsilon_{2} (t)$ for $|\varepsilon_{1} (t)|<|\varepsilon_{2} (t)|$ and $\dot{\varepsilon}_{1}  (t)=-\varepsilon_{1} (t)$ for $|\varepsilon_{1} (t)|\geq|\varepsilon_{2} (t)|$. Note that $\dot{\varepsilon}_{2}  (t)=-\varepsilon_{2} (t)$ for $|\varepsilon_{1} (t)|<|\varepsilon_{2} (t)|$ has no equilibrium point since $\varepsilon_{2} (t)=0$ results in $\varepsilon_{1} (t)=0$, while in this case, $\dot{\varepsilon}_{1} (t)=-\varepsilon_{1} (t)$ for $|\varepsilon_{1} (t)|\geq|\varepsilon_{2} (t)|$ gets activated. Therefore, $\varepsilon_{1} (t)=0$ can be considered as an equilibrium and $|\varepsilon_{2} (t)|=0$ as a state constraint. Note that the state constraint may impose a hard constraint on the original state space variables. Considering $\varepsilon_{2} (t)=0$ and $\varepsilon_{1} (t)=0$, we have the following:

\begin{equation}
\dot{x} _{i} (t)=v_{d} \big(x_{i} (t)\big),
\label{eq6}
\end{equation}
\begin{equation}
x_{i-1} (t)-x_{i} (t)=T\dot{x} _{i} (t).
\label{eq7}
\end{equation}
Therefore, macroscopic parameters of flow $q$ and density $k$, corresponding to the system state at the equilibrium point can be calculated as follows:

\begin{equation}
q(l)=\frac{1}{T},
\label{eq8}
\end{equation}
\begin{equation}
k(l)=\frac{1}{Tv_{d} (l)},
\label{eq9}
\end{equation}
where $l$ denotes the measurement location and $q$, $ k$, and $v$ denote flow rate, density, and velocity, respectively. Two types of fundamental equations can be defined for vehicular traffic: i) conservation of mass \cite{R25} and ii) Wardrop principal for the steady state condition \cite{R26}. The following equations show the mass conservation and Wardrop Principal, respectively.
\begin{equation}
\frac{\partial k(t,x)}{\partial t}+\frac{\partial q(t,x)}{\partial x}=0,\hspace{0.2cm} \forall t,x \in \Bbb R,
\label{eq10}
\end{equation}
\begin{equation}
q(l)=k(l)v_{d} (l).
\label{eq11}
\end{equation}
Note that (\ref{eq8}), (\ref{eq9}) can be utilized in (\ref{eq10}) and (\ref{eq11}) to model and predict traffic flow dynamics. Therefore, the presented approach can be utilized to further manage and regulate the traffic flow. As a result, specific densities and flows could be guaranteed at the steady state condition along with increasing travel time reliability and ensuring safety. This will be discussed in detail in Section \ref{sec:Discussion}.
\end{rem}

\section{Simulation Results}
\label{sec:Simulation Results}
\begin{figure}

    \hspace{0.2cm}\begin{subfigure}{0cm}
        \includegraphics[width=8cm]{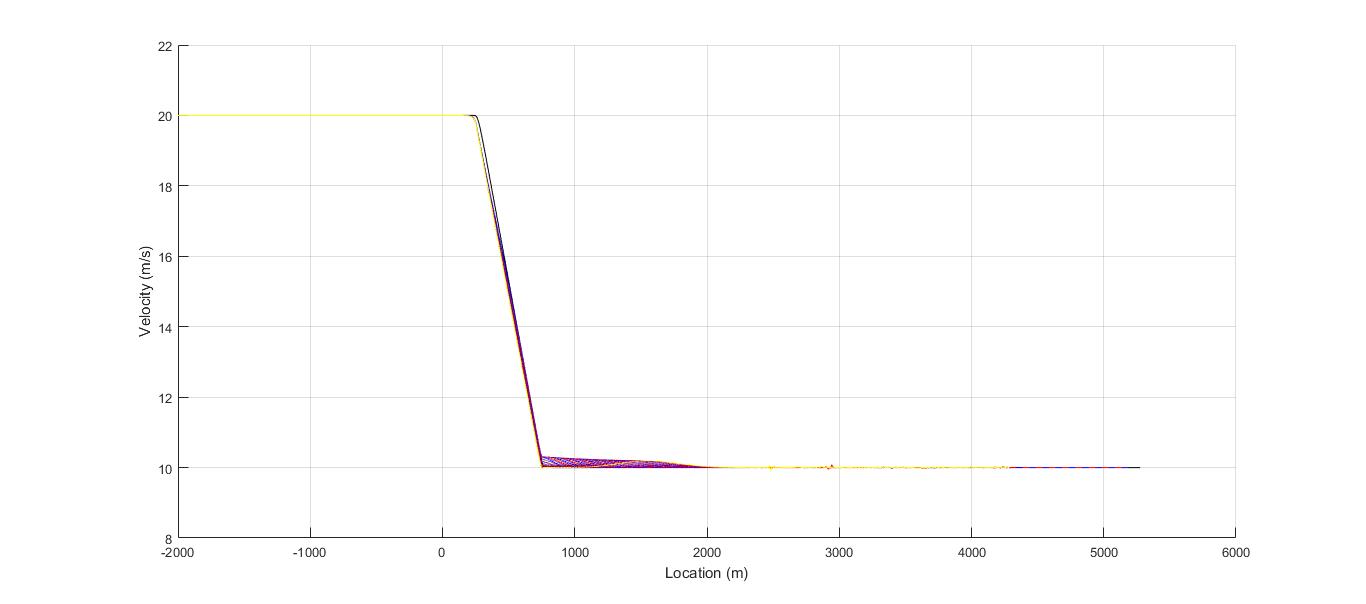}
        \caption{}
           \label{fig2.1}
    \end{subfigure}\\
    \vspace{0cm}
    \begin{subfigure}{0cm}
        \includegraphics[width=8cm]{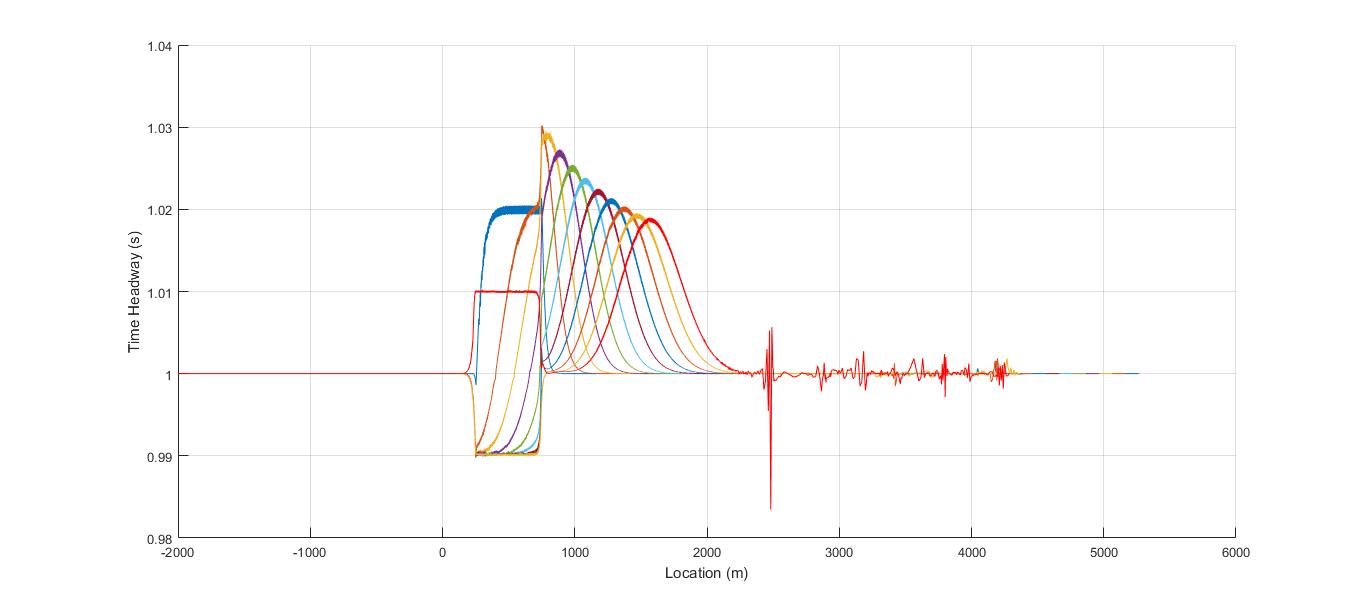}
        \caption{}
           \label{fig2.2}
    \end{subfigure}
    \caption{Simulation results for Scenario 1, where vehicles try to follow the velocity profile without any external perturbation: (a) velocity profile for vehicles 10, 20, 30, 40, 50, 60, 70, 80, 90, and 100, and (b) time headway evolution over distance for the same vehicles. In Figure \ref{fig2.1} yellow line indicates velocity profile of the last vehicle. The red line in Figure \ref{fig2.2}  is also a representative of Time Headway for the last vehicle. }
\label{fig2}
\end{figure}

\begin{figure}

    \hspace{0.2cm}\begin{subfigure}{0cm}
        \includegraphics[width=8cm]{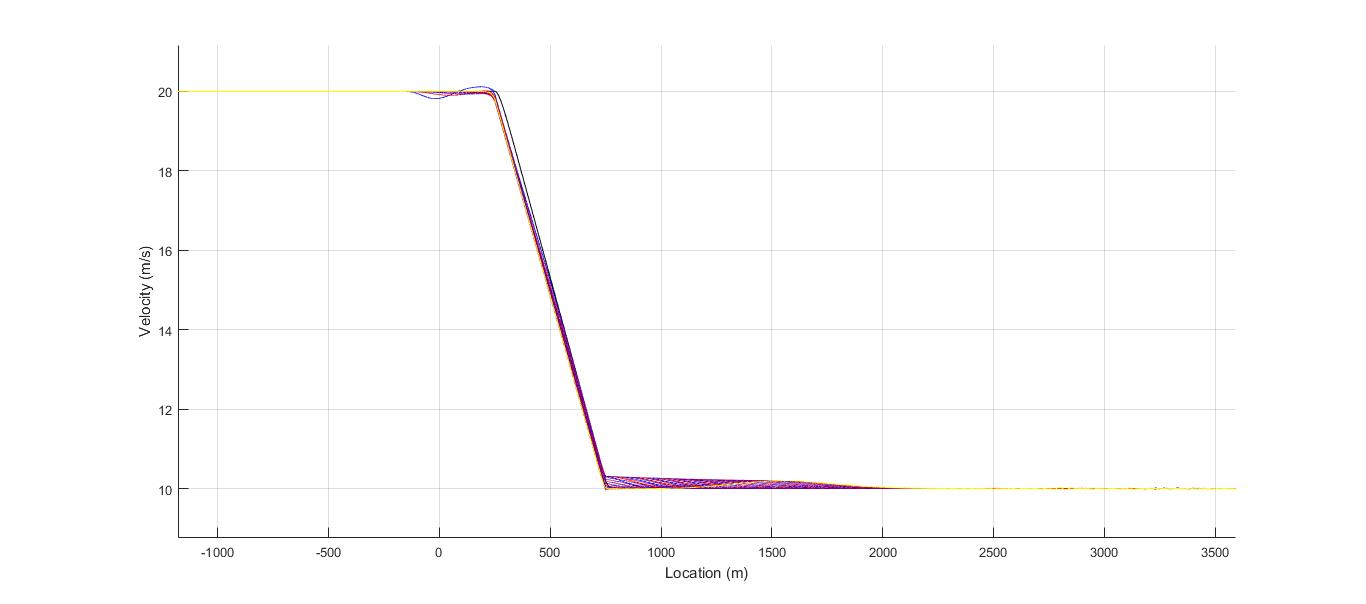}
        \caption{}
           \label{fig3.1}
    \end{subfigure}\\
    \vspace{0cm}
    \begin{subfigure}{0cm}
        \includegraphics[width=8cm]{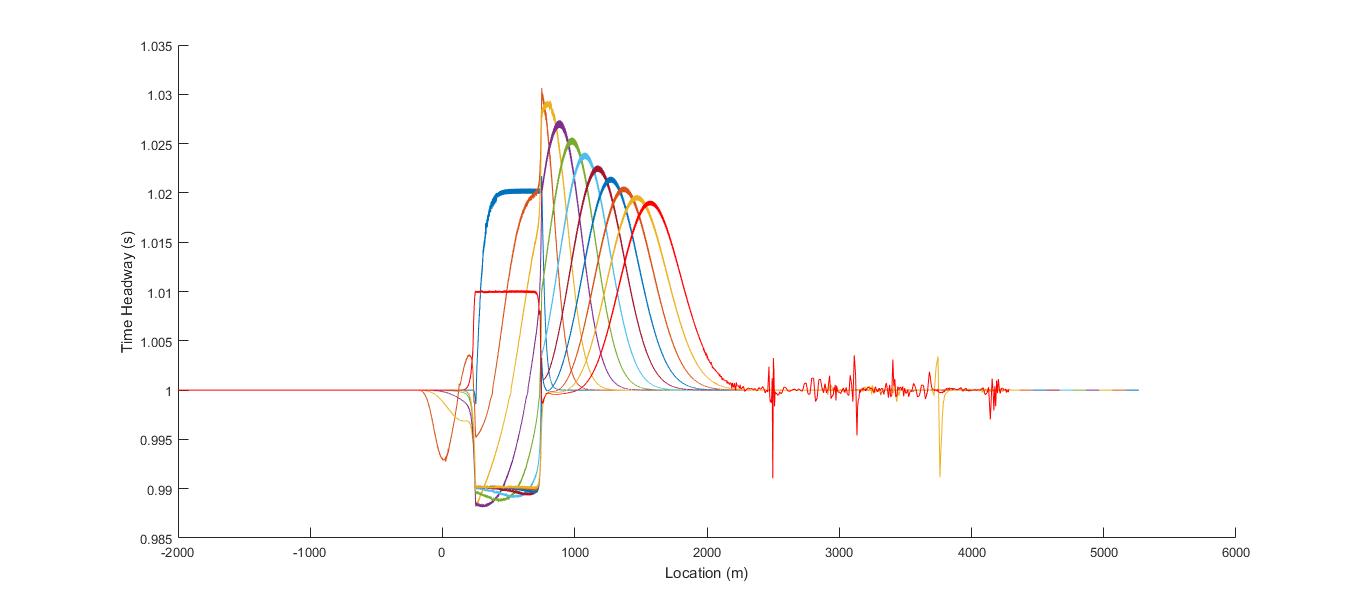}
        \caption{}
           \label{fig3.2}
    \end{subfigure}
    \caption{Simulation results for Scenario 2, where vehicle 3 is perturbed for 10m: (a) velocity profile for vehicles 10, 20, 30, 40, 50, 60, 70, 80, 90, and 100, and (b) time headway evolution over distance for the same vehicles. In Figure \ref{fig3.1} yellow line indicates velocity profile of the last vehicle. The red line in Figure \ref{fig3.2} is also a representative of Time Headway for the last vehicle.}
\label{fig3}
\end{figure}

In the previous section, we showed that the proposed controller is capable of following a desired speed profile, while ensuring safety. In this section, we investigate the behavior of the proposed controller through a series of simulations. Accordingly, we simulate the dynamic system of the form (\ref{eq5}) for a gradual speed drop from 20m/s to 10m/s for time headway of one second. The length of the speed drop is 500m. A string of 100 vehicles is simulated for two scenarios. In scenario 1, no error in the initial condition is considered, while in scenario 2, the location of vehicle 3 is perturbed for 10m. Figures \ref{fig2} and \ref{fig3} present the results of these simulations. Figure \ref{fig2.1} shows the  velocity profile versus location for vehicles 10, 20, 30, 40, 50, 60, 70, 80, 90, and 100 for Scenario 1, while Figure \ref{fig2.2} presents the changes in time headway over distance for the above vehicles. Figures \ref{fig3.1} and \ref{fig3.2} shows the similar graphs for Scenario 2.

Based on Figures \ref{fig2.1} and \ref{fig3.1}, the proposed system accurately follows the velocity profile for both cases (with and without external perturbation). Figures \ref{fig2.2} and \ref{fig3.2} indicate that the calculated time headway only fluctuates within a small bound (i.e., [0.98,1.04]), which indicates that the system ensures safety. Note that these small fluctuations are due to discretization by the solver (Matlab's ODE113 solver is used to solve the system of differential equations).
Overall, velocity profiles of all vehicles are similar to the desired velocity profile given in Figure \ref{fig1} (with small deviations from the desired profile for the first few vehicles). Moreover, external perturbations disappear such that, even in Scenario 2, the velocity profile of vehicle 100 (yellow line) is almost identical to the desired velocity profile and its time headway reaches 1s.

\section{Discussion}
\label{sec:Discussion}
In previous works by Besselink and Johanson \cite{R13} and Shaw et al. \cite{R27}, controllers in space and time domains were proposed to ensure string stability (and bounded errors, consequently), while tracking the velocity profile of a leading vehicle with constant time headway and constant spacing policies, respectively. However, their methodology requires information about other vehicles movement (particularly, they either need information about the time that each vehicle is at a certain location or location of other vehicles at a particular time). The safety and efficiency of these methodologies can be influenced by the availability of the information, which should be transmitted to all vehicles in the platoon using V2X communications. As discussed previously, due to several factors, including signal interference, such information might not be available at all times. Therefore, those and similar approaches might not apply to the real-world applications. Moreover, the performance of these methodologies can degrade with increasing the number of vehicles in the platoon and/or uncertainties in communications. Moreover, in those studies, control laws are based on the time delay in the presence of speed profile. Therefore, achieving a particular flow in the link may not be guaranteed as the flow is defined base on time headway, which is not under control. Additionally, maintaining a constant time headway is a reliable measure to ensure safety, while a time delayed spacing policy may not be as effective. In the current research, we proposed a decentralized controller based on manipulating maximum error associated with a vehicle movement (error in velocity concerning the velocity profile and error in the location with respect to the leading vehicle). Assuming that full information of states is accessible without an observer, sufficient conditions to avoid the collision and keeping acceleration/deceleration bounded are straight forward to obtain based on the proposed method . Analysis showed that the proposed control method guarantees the string stability. Mathematical analyses presented in the paper are valid for any velocity profile that satisfies the conditions expressed in the Theorem, such as being Lipschitz with Lipschitz constant $M$ satisfying $TM<1$, where T is the constant Time Headway.

Overall, the main advantage of the proposed control method over the previously discussed studies is the decentralized nature of it. Such a characteristic provides robustness against uncertainties in the communications network as the controller only needs information about the leader's location and speed. Such information is available through on-board sensors (e.g., radar and LiDAR).

\section{Conclusion}
\label{sec:Conclusion}
Connected and autonomous vehicles continue to promise improvements in safety and efficiency. Even though the current practice of designing and operating autonomous vehicles is based on isolated vehicles, platooning is expected to significantly enhance the efficiency throughout the transportation system (e.g., improving congestion and reducing emissions and energy consumption). There are, however, particular challenges that need to be addressed to ensure a safe and reliable operation of the platoons of autonomous vehicles. One of the key challenges is following a certain velocity profile (e.g., changes in speed limit due to work zone operations), while ensuring safety. While very few studies focused on this problem, they rely on accurate and timely information from V2V and V2I communications network. Such information is not always available, which can cause efficiency and possibly safety issues. Accordingly, this study, using the state feedback of the form (\ref{eq1}) for the simple vehicle dynamics of the form $\ddot{x} _i (t)=u_i$, presents a methodology that, for some desirable initial conditions, avoids collision throughout the time and space, while tracking a desired velocity profile. Tracking the desired velocity profile ensures high travel time reliability. Additionally, predetermined constant time headway ensures predictable traffic flow dynamics, which is important from a system perspective and can be utilized for congestion prediction and mitigation. We showed that error terms corresponding to the velocity profile and safe distance are adjusted asymptotically. It is noteworthy that using the state feedback loop, all error terms remain bounded throughout the time and space. Our simulation results also indicated that the proposed methodology is robust to internal and external disturbances in the system and can follow the desired profile, while ensuring safety. This method could be potentially useful to stabilize nonlinear interconnected systems, where number of outputs are greater than number of inputs. Accordingly, generalizing this approach has been left for future research.

\bibliographystyle{IEEEtran}
\bibliography{RefDataBase}

\vspace{-30 mm}
\begin{IEEEbiography}[{\includegraphics[width=1in,height=1.25in,clip,keepaspectratio]{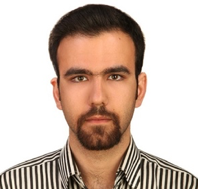}}]{Sina Arefizadeh} is currently a PhD student of Transportation Engineering
at Texas A\&M University. He received his M.Sc. in Civil Engineering from Sharif University of Technology in 2011 and his B.S. degree in Civil Engineering from Ferdowsi university in 2013. His research interests includes: linear and non-linear control systems, Dynamic Programming, Traffic Modeling and
Simulation, Intelligent Transportation System, Mathematical Analysis,
and Queuing Theory.
\end{IEEEbiography}
\vspace{-30 mm}
\begin{IEEEbiography}[{\includegraphics[width=1in,height=1.25in,clip,keepaspectratio]{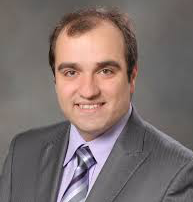}}]{Alireza Talebpour} received his B.S. and M.S. degrees in Civil Engineering from Sharif University of Technology,
Tehran, Iran, in 2007 and 2009, respectively. He received his Ph.D. in Civil and
Environmental Engineering from Northwestern University, Evanston, IL,
USA, in 2015. He is currently an Assistant Professor in the
Zachry Department of Civil Engineering at Texas A\&M
University, College Station, TX, USA. He has pursued multiple research
topics, including driver behavior modeling, traffic flow theory, safety, and
intelligent transportation systems. His current research focuses on
connected and automated transportation systems.
\end{IEEEbiography}
\vspace{-30 mm}
\begin{IEEEbiography}[{\includegraphics[width=1in,height=1.25in,clip,keepaspectratio]{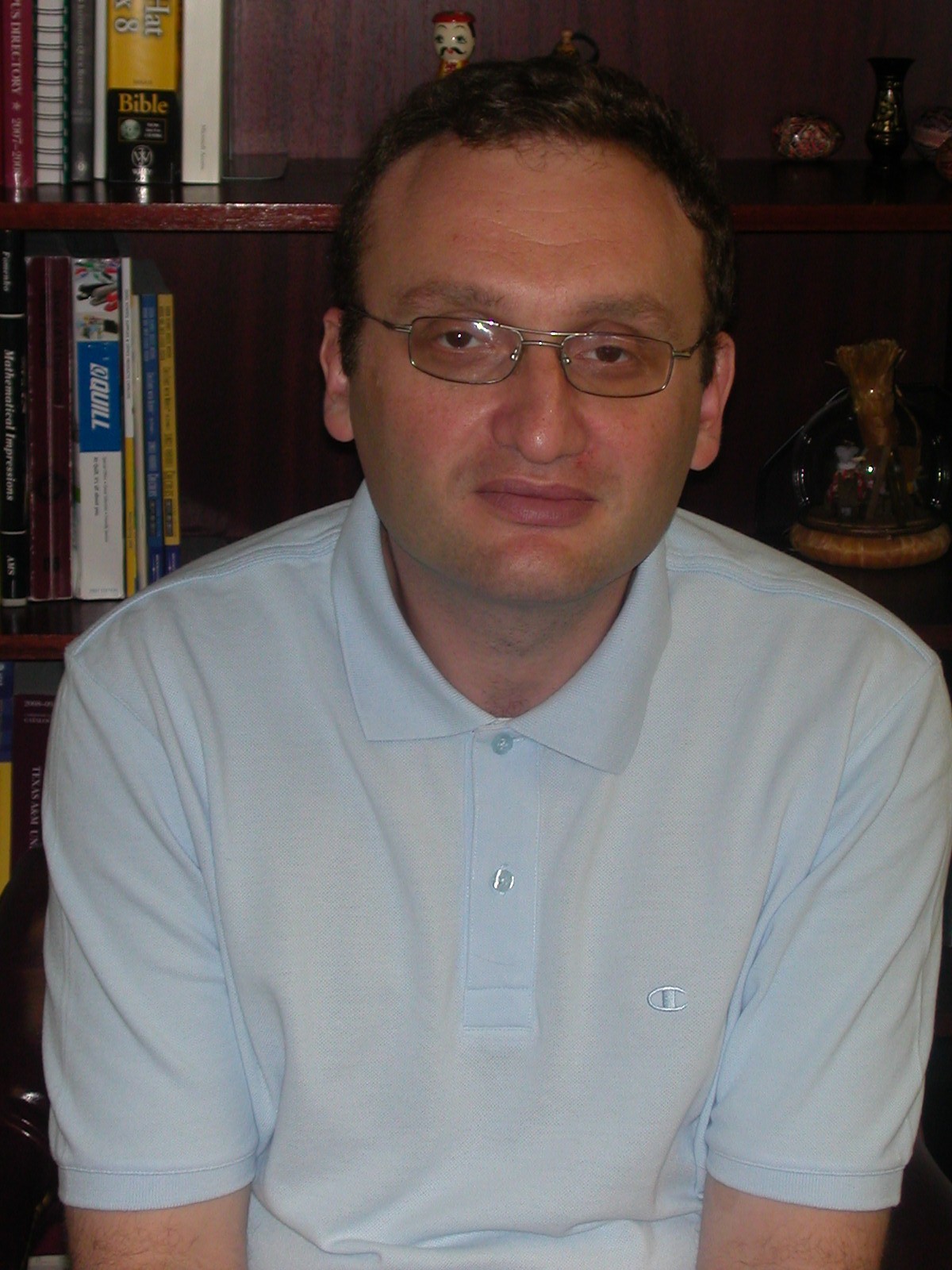}}]{Igor Zelenko} received his B.S., M.S., and PhD degrees in Mathematics from Technion- Israel Institute of Technology, in 1994, 1998, and 2002 respectively. He is currently an Associate Professor in the
Department of Mathematics at Texas A\&M University, College Station, TX, USA. He was a Research Associate and Visiting Associate Professor at International School for Advanced Studies (SISSA-ISAS), Trieste, Italy. He won several national and international grants including NSF and Simons Collaboration Grants and Research Grants from Italian Ministry of Education and Science (MIUR). He is currently active in Differential Geometry and Geometric Control Theory. His main direction of current research is the construction of canonical frames and differential invariants for a wide class of geometric structures and control systems on manifolds.
\end{IEEEbiography}

\end{document}